\documentclass{iucr}

\usepackage{graphicx}
\usepackage{xspace}
\usepackage{xcolor}

     \journalcode{B}              

\usepackage{amsmath}
\usepackage{bm}

\usepackage{braket}

\newcommand{\mm}[1]     {\ifmmode {#1} \else{}${#1}$\fi}
\newcommand{\mmm}[1]    {\ifmmode{}#1 \else{}${#1}$\fi}
\newcommand{\beq}[1]{\begin{equation}\label{#1}}
\newcommand{\eeq}{\end{equation}}

\def \tbag{\mm{\rm Tb_{14}Ag_{51}}}

\def\vec#1{\mm{{\rm\bm{{\mathrm#1}}}}}

\def \oth{{\mm{{1\over3}}}}

\def\kstar{\mm{ \pm[\oth,\oth,0]}}
\def\onek{\mm{[\oth,\oth,0]}}

\def\Psix{\mm{P\bar{6}'}}
\def\Pm{\mm{P{m}'}}
\def\P6m{\mm{P{6}/m}}

\def\figsiz{\columnwidth}

\begin{document}                  



\title{Revisiting the antiferromagnetic structure of \tbag . The importance of distinguishing alternative symmetries for a multidimensional order parameter} 
\shorttitle{Antiferromagnetic structure of \tbag}


\author[a]{V.}{Pomjakushin}  
\author[b]{J.M.}{Perez-Mato} 
\author[a]{P.}{Fischer} %
\author[a]{L.}{Keller} %
\author[c]{W.}{Sikora} %
\aff[a]{Laboratory for Neutron Scattering and Imaging (LNS), Paul Scherrer
Institut, CH-5232 Villigen PSI \country{Switzerland}}
\aff[b]{Facultad de Ciencia y Tecnolog\'{i}a, Universidad del Pa\'{i}s Vasco, UPV/EHU, Apartado 644, E-48080 Bilbao \country{Spain}}
\aff[c] {Faculty of Physics and Applied Computer Science, AGH University of Science and Technology, PL-30-059 Krakow, \country{Poland}}








\maketitle                        
\begin{synopsis}
The magnetic structure of Tb14Ag51 is newly modelled under a magnetic space group, which is one of the possible alternative symmetries for the relevant irreducible representation. This approach results in a model that fits the data much better than the one previously reported.
\end{synopsis}


\begin{abstract}

We revisit the antiferromagnetic structure of \tbag\ \cite{ISI:000232933700063} with the propagation vector \onek\ and parent space group \P6m\ using both magnetic symmetry and irreducible representation arguments. We have found a new magnetic structure under the hexagonal Shubnikov magnetic space group \Psix, which fits much better the experimental data. This new solution was obtained by constraining the spin arrangement to one of the three possible magnetic space groups of maximal symmetry that can be realised by a magnetic ordering transforming according to the 4-dimensional physically irreducible representation that is known to be relevant in this magnetic phase. The refined model, parameterised under  \Psix, implicitly includes the presence of a third harmonic with the propagation vector at the gamma point $[0,0,0]$, which has an important weight in the final result. The structure consists of 13 symmetry-independent Tb magnetic moments  with the same size of  $8.48(2)\mu_B$, propagating cycloidally in the ab-plane. The modulation has a substantial deviation from being purely sinusoidal due to the contribution of the mentioned third harmonic.

\end{abstract}


\section{Introduction}

Intermetallic uranium and rare-earth $\rm A_{14}B_{51}$ compounds with $\rm Gd_{14}Ag_{51}$ structure have interesting physical properties such as coexistence of antiferromagnetic order and heavy-fermion behaviour in $\rm Ce_{14}X_{51}$ (X = Au, Ag, Cu) and in $\rm U_{14}Au_{51}$. This is related to the fact that there are three crystallographically distinct A sites in this structure, implying considerable geometric frustrations. 
The interesting magnetic properties of these compounds motivated the  powder neutron diffraction investigations of the similar antiferromagnetic rare-earth compound \tbag\  \cite{ISI:000232933700063} in order to determine to which extent its magnetic ordering is different, and in particular to investigate whether there can also be magnetic order on all rare-earth sites. 
The above study has shown that the magnetic structure has propagation vector  \vec{k} = \onek\  (K-point of the Brillouin zone) in the parent space group \P6m\ with three magnetic Tb sublattices ordering simultaneously according to the (physically) irreducible representation (irrep) mK4K6, corresponding to Shubnikov magnetic space group (MSG) \Pm, below $T_N$ = 27.5(5)~K (we use here internationally established nomenclature for the irrep labels  and MSG \cite{Bilbao,isod}). In the above mentioned study the best solution for the model with only a K-point magnetic modulation has been found, but due to the fact that no symmetry arguments were used a better  solution of higher symmetry was missed.  
Being 4-dimensional, the irrep of the magnetic order parameter (OP) in fact allows the realisation in the magnetic phase of different alternative MSGs depending on its direction. The model proposed in \cite{ISI:000232933700063} corresponds to a general OP direction with the lowest possible symmetry. 

Our motivation to revisit this structure was the fact that possible alternative models of higher symmetry compatible with the identified active irrep had not been considered, and in fact most of the spins in the reported model seemed to comply with one of these possible higher symmetries, namely the MSG \Psix. Only the spins generated for Tb2 in orbit 1, according to the description in \cite{ISI:000232933700063} do not fit into a \Psix\ model, and  break the symmetry into the lowest possible  one for this irrep mK4K6: \Pm. This means that 99 spins in the 3x3x1 cell of the published model comply with the MSG \Psix, while 27 do not. We therefore undertook a new refinement under a model fully consistent with the MSG \Psix. This model not only automatically constrains the degrees of freedom associated with the active irrep mK4K6 to comply with the assumed MSG, but it also implicitly includes all spin degrees of freedom with $k=0$, which are necessarily allowed under this symmetry. This secondary modes can be considered as third harmonic with respect to the primary mK4K6 modulation. In this article we present the  magnetic structure obtained under this approach, which fits the experimental data much better than the original model proposed in \cite{ISI:000232933700063}. 

\section{Experimental}

The experimental data which are used in the present paper are the same that were used in the previous paper \cite{ISI:000232933700063}.
These data were the result of neutron powder diffraction measurements carried out at the high-resolution HRPT 
diffractometer (High-Resolution Powder Diffractometer for Thermal Neutrons) \cite{hrpt} 
and DMC   
(Double axis Monochromatic Cold diffractometer)
at the SINQ neutron spallation source (PSI, Switzerland). 
The high intensity mode and neutron wavelengths $\lambda\!=\!1.886$~\AA\ and $\lambda\!=\!2.567$~\AA\ were used, respectively on these instruments. The crystal structure parameters were determined from a HRPT diffraction pattern at T=30~K, and then were fixed in the magnetic structure solution from the DMC data. We used a difference pattern between 1.5~K and 30~K diffraction patterns measured at DMC, which contains a purely magnetic contribution, assuming that there are no significant structure changes between these two low temperatures.
To further prove that we indeed do not have the significant changes in the atomic positions we have undertaken a combined refinement of HRPT and DMC patterns at T=1.5~K. 
The refinements of the structure parameters were done using the {\tt FULLPROF}~\cite{Fullprof} program, with the use of its internal tables for neutron scattering lengths. 
The symmetry analysis was performed using {\tt ISODISTORT} from the {\tt ISOTROPY} software \cite{isod,isod2}, some software tools of the Bilbao crystallographic server as  {\tt k-SUBGROUPSMAG} and  {\tt MAXMAGN} \cite{Bilbao,ISI:000358484200010} and the {\tt BasiReps} program \cite{Fullprof}. The plots of magnetic structures were made with the help of  {\tt VESTA} software \cite{VESTA}.

\section{Symmetry analysis}

The space group of the paramagnetic crystal structure of \tbag\ is \P6m\ with  Tb-atoms in  the following positions: Tb1 6k (x,y,1/2), Tb2 6j (x,y,0) and  Tb3 2e (0,0,z). The primary propagation vector is $\vec{k}_K$ = \onek, labeled as K-point of the Brillouin zone. 
The active irrep mK4K6 that was identified in \cite{ISI:000232933700063} is the result of the direct sum of two mathematically irreducible representations that are complex conjugate (mK4 and mK6). As the small irrep for each of them is 1-dimensional and $k$ and $-k$ are not equivalent, the resulting real physically irreducible representation is 4-dimensional (the m in the label indicates that the irrep is odd for time reversal to be distinguished from the irreps associated with non-magnetic degrees of freedom as phonon modes, which are even). 

Figure 1 depicts the possible MSGs that can be realised by an OP transforming according to the 4-dimensional irrep mK4K6, as obtained with the program  {\tt k-SUBGROUPSMAG} of the Bilbao Crystallographic Server. One can see that there are three possible maximal symmetries for special OP directions (irrep epikernels), while for a general direction, i.e. for an unconstrained combination of the irrep spin basis functions, the symmetry is reduced to the MSG \Pm\  (irrep kernel). The structure proposed in \cite{ISI:000232933700063} did not assume any particular constraint on the OP direction and therefore the model only has the lowest possible symmetry. This is given by the MSG \Pm\ $(2a+b,c,a-b; 0,0,0)$ (in parenthesis a basis transformation to the standard setting of the MSG, which fully defines the symmetry group, is indicated). This MSG, however, allows third harmonic secondary modes according to irreps at the gamma point (GM) $k=0$, which were not considered. 
This was the logical consequence of the use of the traditional irrep-approach, where magnetic symmetry is disregarded.

\begin{figure} 
\begin{center}
\includegraphics[width=\figsiz]{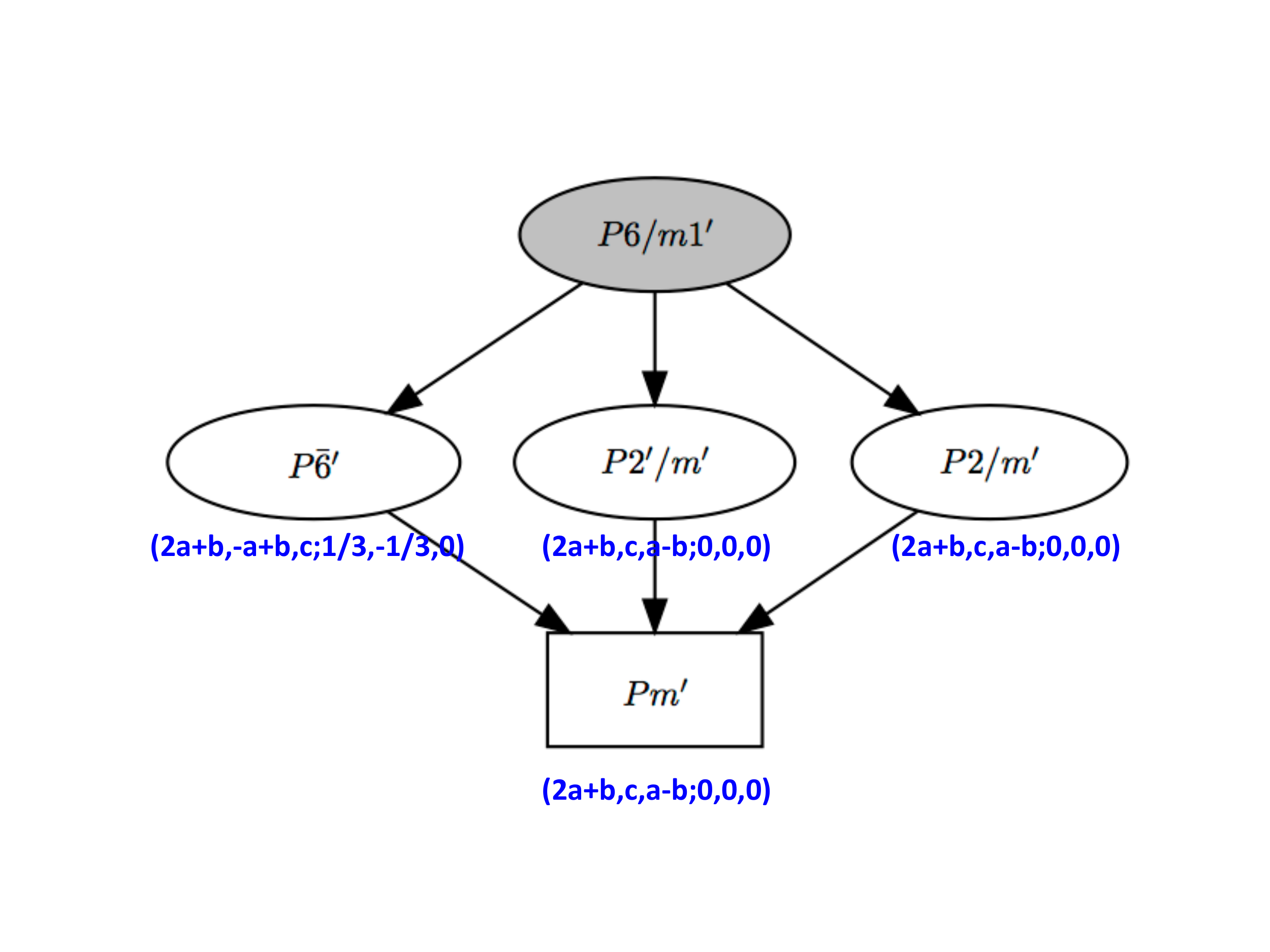} 
\end{center}
\caption{Possible alternative magnetic symmetries that can be realised as the result of a magnetic ordering on a \P6m\  paramagnetic structure, if the spin arrangement transforms according to the 4-dimensional physically irreducible representation mK4K6. They are depicted in a group-subgroup hierarchical form, as subgroups of the grey MSG \P6m{1'} of the paramagnetic phase. Only one subgroup for each set of equivalent conjugate subgroups is shown and it is fully defined by a basis transformation that would put the subgroup in the MSG standard setting (obtained with k-SUBGROUPSMAG \cite{ISI:000358484200010}).
}
\label{subgroups}
\end{figure}

Modelling the magnetic structure under one of the MSGs represented in Figure \ref{subgroups}, without any further restriction, not only allows the presence of spin modes according to the irrep mK4K6 (constrained in a certain form if the MSG is not the irrep kernel), but it also automatically includes the secondary spin degrees of freedom corresponding to $k=0$, which are symmetry allowed as a third harmonic of the primary propagation vector \onek. This is a general feature when modelling under a MSG: while the assumption of one particular MSG may constrain the possible combinations of irrep basis functions for the primary irrep, and therefore it can reduce the number of free parameters for the primary spin arrangement, it also automatically includes all magnetic degrees of freedom corresponding to secondary irreps, which are symmetry allowed. For instance, odd harmonics will be present in the MSG model if their wave vector is not equivalent to the primary one. This can often be a nuisance, as these high-order secondary spin modes may considerably increase the number of degrees of freedom of the structure, when in fact they can often be neglected. But in the present case, on the contrary, the weight of a secondary $k=0$ component is in fact, as shown below, very important, and a model using only the constraints of the MSG, results to be an optimal approach to take into account its presence in the structure.

For the reasons explained in the introduction, we undertook the refinement of the magnetic structure under the constraints of the MSG \Psix\ (BNS index 174.135 ) shown in Figure 1. For simplicity, the structure was transformed to the standard setting of this MSG.  Its unit cell volume is then 3 times larger than in the parent cell, with a basis transformation (1,-1,0),(1,2,0),(0,0,1) and origin shift $(1/3,2/3,0)$, which is an equally valid transformation as the one in Figure 1, reproduced from the  {\tt k-SUBGROUPSMAG} output. Under this symmetry group the Tb1 and Tb2 sites in the paramagnetic phase split into six 3k and six 3j sites, respectively, which only allow non-zero magnetic moments on the basal plane, while the Tb3 site remains a single site, but in a general position 6l of the MSG, and therefore with no constraint on its magnetic moment. This means that the model has 27 free parameters to describe the spin arrangement, which reduce to 15 if equal modulus of all magnetic moments is assumed.   

\subsection{A note on the relations between irreps and magnetic symmetry}
In a general case, the magnetic degrees of freedom that can be spontaneous as secondary effects resulting from the coupling with the primary order parameter are all those that can couple linearly with some power of the primary order parameter (a polynomial of the order parameter components). In this respect  the secondary irreps should not transform in the same manner as the primary irrep. They should transform in the same way as some polynomial $ P_n(S_1,...,S_n)$ of some power n, so that their product is an invariant. To derive then these possible secondary irreps is therefore not trivial. One can derive automatically that secondary irreps with $k'=3k$  are possible, if $3k$ is not equivalent to $k$, because then invariants between a polynomial of third power of the order parameter and a secondary magnetic order parameter with wave vector $3k$ will be possible, but which particular irreps for $3k$ are those having these invariants is not known and has to be calculated ( {\tt ISODISTORT} does it in one of its side options).

The point is that once one has assigned one of the possible MSGs resulting from the primary irrep, then the degrees of freedom within this MSG will include  automatically those resulting from these possible linearly coupled secondary irreps. This is an advantage  but if the description is done only using the MSG (without modes), in an atomistic way, as we did, the degrees of freedom corresponding to the primary irrep and the secondary ones are ``entangled'' and add up. This was no problem in our case,  but in many cases this may be a disadvantage, because the secondary modes may be negligible, and could be made zero, decreasing the number of parameters. That is why in the case of a MSG allowing more than one irrep, the best approach is to use both the MSG and also a decomposition into irrep basis functions but  restricted to this MSG, as done in  {\tt ISODISTORT}, and in the  {\tt FULLPROF} option using the output of  {\tt ISODISTORT}.



\section{Powder diffraction data analysis}

\subsection{Model free Le Bail fit}
\begin{figure} 
\begin{center}
\includegraphics[width=\figsiz]{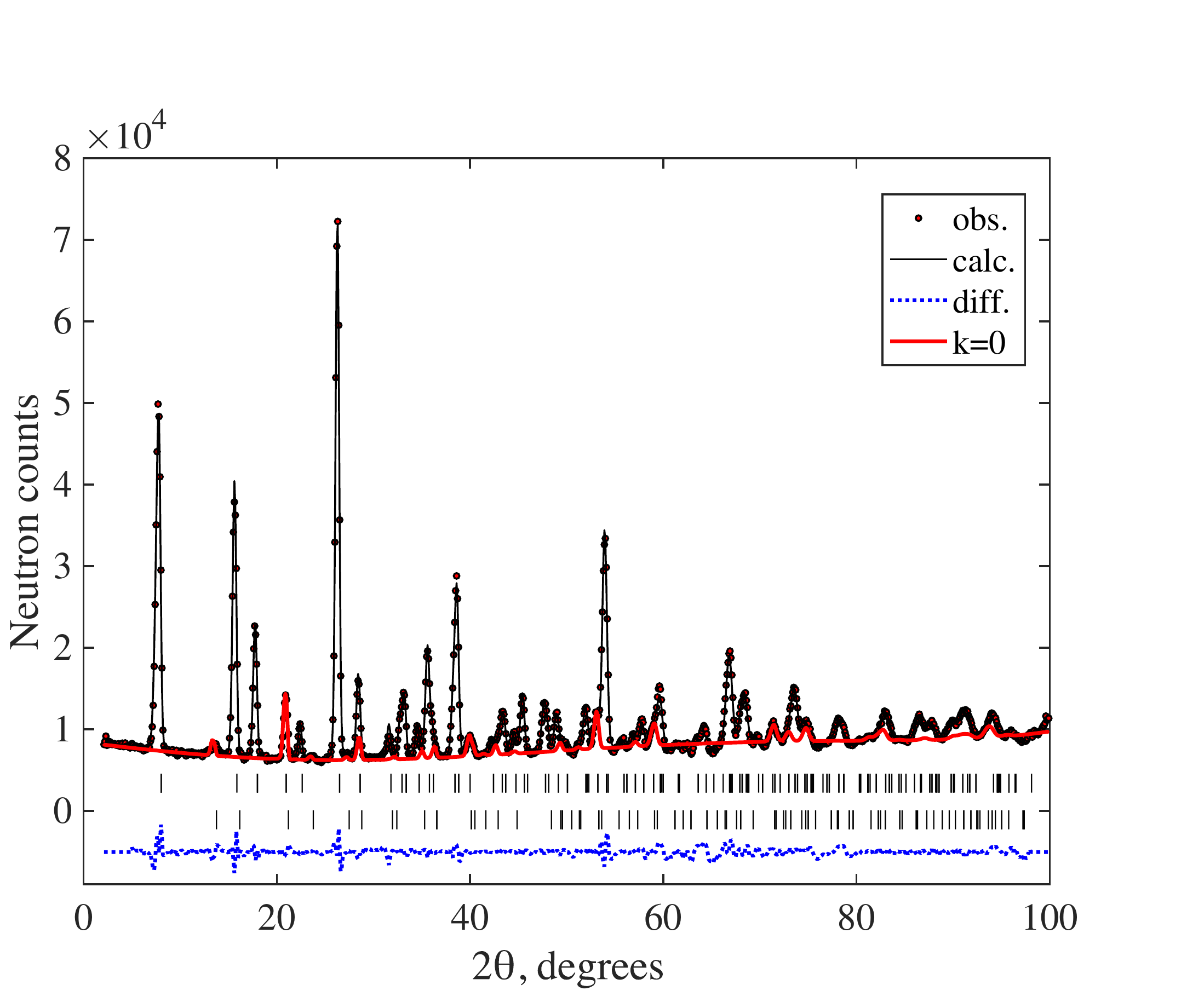} 
\end{center}
\caption{Neutron powder diffraction pattern showing the difference between data measured at $T=1.5$\,K and $T=30$\,K, with wavelength $\lambda=2.567$\,\AA. The solid line shows result of the Le Bail fit. The rows of vertical tick mark the hkl's of the Bragg peaks for propagation vectors \kstar (upper) and [0,0,0]. The contribution of [0,0,0] is shown by the red curve. The reliability factors~\cite{Fullprof} are: Rp:  3.14, Rwp:  4.19, Rexp: 1.29, Chi2: 10.6.}
\label{lebail}
\end{figure}

A first step in the determination of a magnetic structure is the identification of the propagation vectors. This is done by the use of the so called LeBail fitting,  where all peak intensities are refined separately without any structure model, thus allowing to fit the propagation vectors and the crystal metrics. In the present case, due to the relatively small primary propagation vector $k_K=\onek$ (K-point) the LeBail fit assuming $k_K$ as the only magnetic wave vector is quite acceptable as shown in the previous paper~\cite{ISI:000232933700063}. However, one can see a noticeable contribution of GM-point $k_0=0$  in quite a few peaks. The fit quality is illustrated in Fig.~\ref{lebail}. For example, the diffraction peaks at $2\theta$ about 13.6 and 21~degrees with quite significant intensity are indexed as (100) and (011), respectively, and therefore indicate a significant presence of a $k=0$ component in the magnetic arrangement. The peak (011) is  close to the peaks (010)/(020)$\pm k_K$, and therefore were indexed as such in the previous study, but a detailed inspection shows that the shift from the K-point positions  is significant. 

The reliability factor for the LeBail fit of difference pattern $\chi^2$ improved from 18.9 to 10.6 when both propagation vectors were used. We note however that the assignments of the experimental intensities to the Bragg peaks from $k_0$ and $k_K$ is not unique and unambiguous in case where the peaks are close in $2\theta$ positions, and the actual evidence of the \Psix\ symmetry will come from the magnetic structure refinement, where the partial contributions to the peak intensities are fixed by the model parameters. The results of this new Le Bail fit are included here in order to define the best possible reliability factor $\chi^2$ that can be attained by the new model. 

\subsection{Structure models}
\begin{figure}
\begin{centering}
\includegraphics[width=\figsiz]{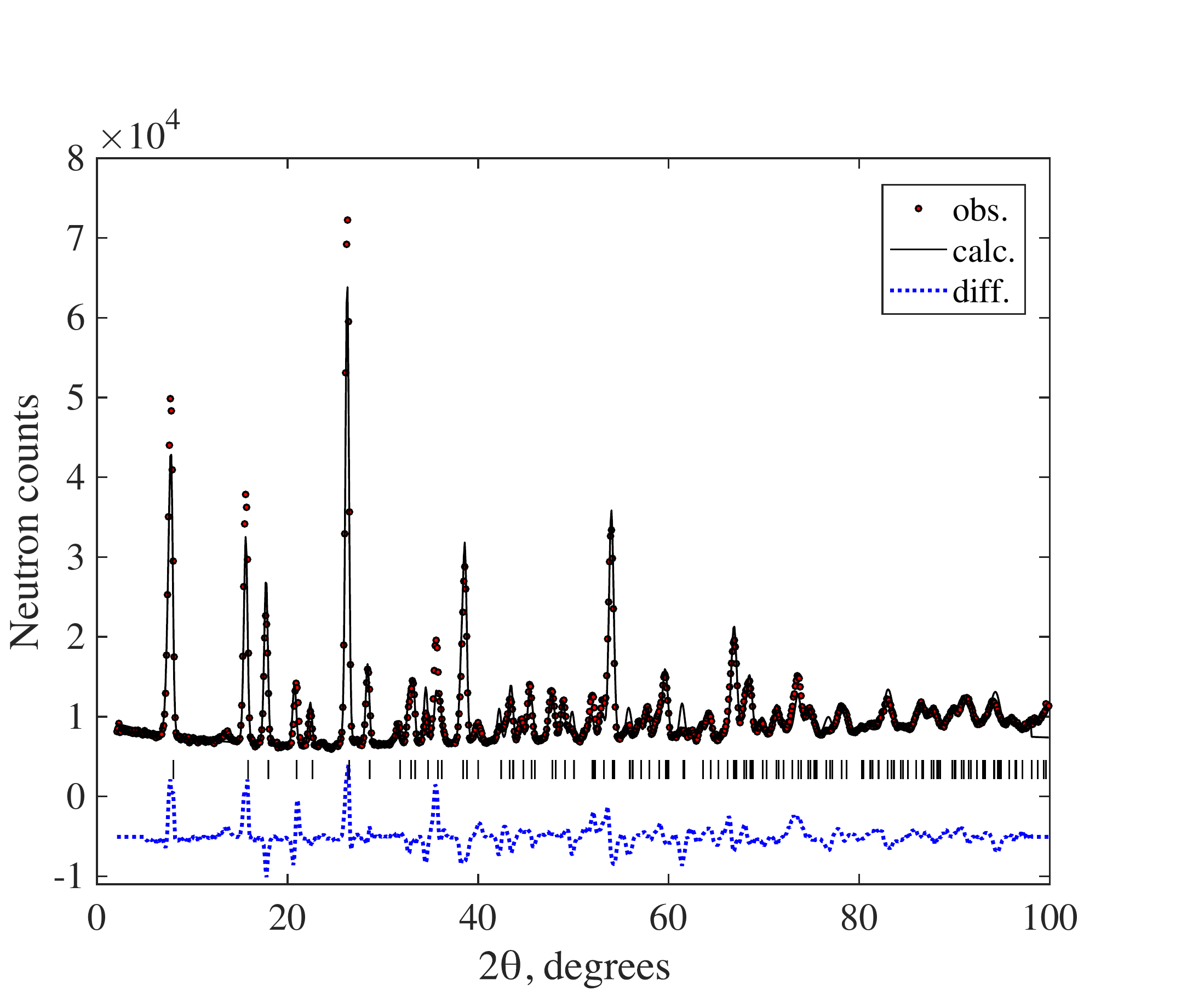} 
\par\end{centering}
\caption{Neutron powder diffraction pattern showing the difference between data measured at $T=1.5$\,K and $T=30$\,K,  with wavelength $\lambda=2.567$\,\AA. The solid line shows the result of the fit to the model used in Ref.~\cite{ISI:000232933700063}. The rows of vertical tick mark the hkl's of the Bragg peaks for the propagation vectors \kstar. The reliability factors are:  Rp:  7.98,  Rwp:  11.2,  Rexp: 1.31, Chi2:  73.0. }
\label{oldK_k}
\end{figure}

\begin{figure}
\begin{centering}
\includegraphics[width=\figsiz]{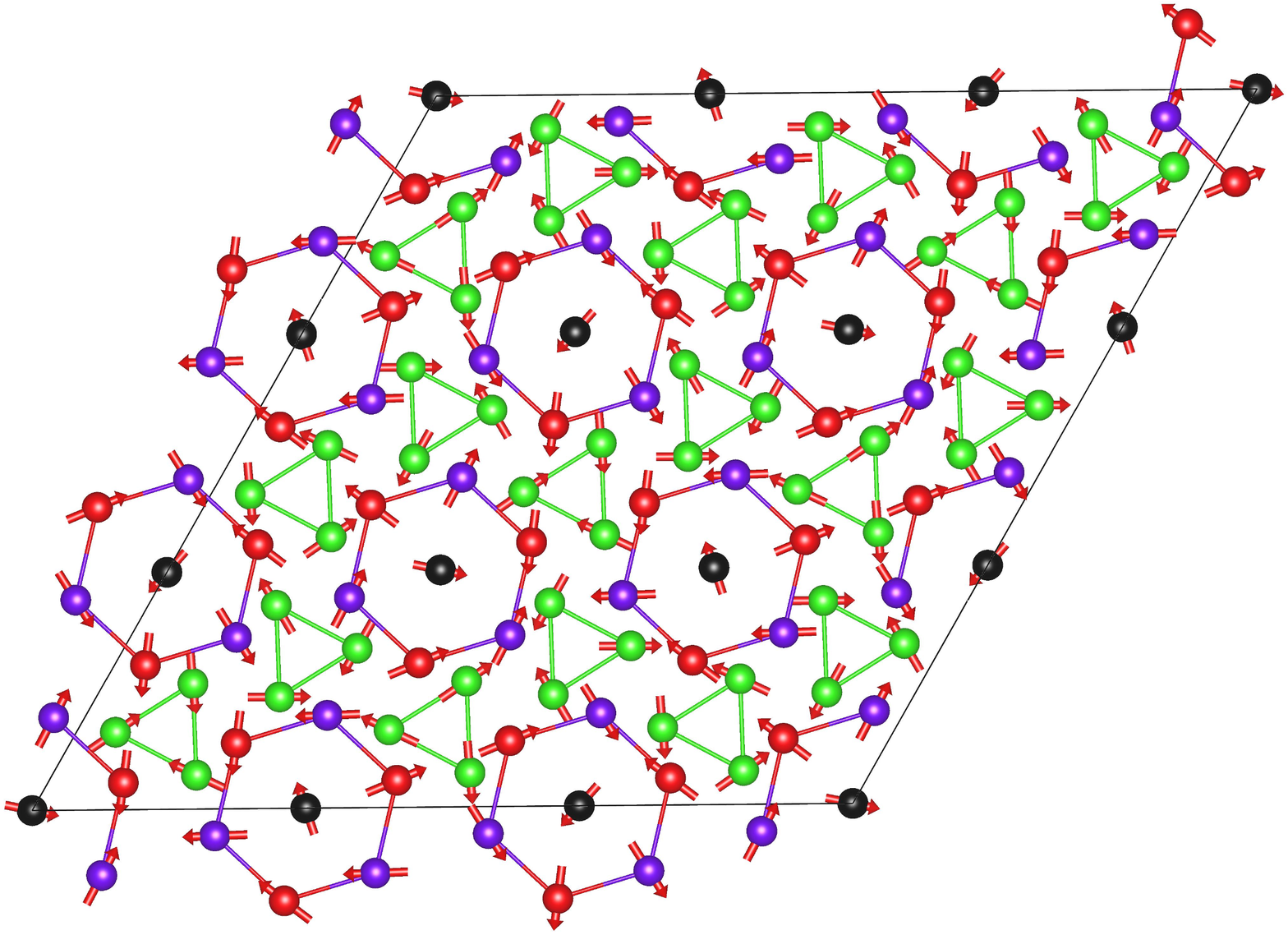} 
\par\end{centering}
\caption{View of the magnetic structure in projection on xy-plane corresponding to the model from Ref.~\cite{ISI:000232933700063}. Tb1 atoms are in green forming the triangles, Tb3 atoms are in black being in the centres of hexagons. Tb2a\_1, Tb2a\_2 and Tb2a\_3 (orbit 1 of Tb2 \cite{ISI:000232933700063}) are in red, and the rest three atoms of Tb2 are in blue. The first three Tb2 atoms reduce the magnetic symmetry down to \Pm.}\label{old_Pm_irreps} 
\end{figure}

\begin{figure}
\begin{centering}
\includegraphics[width=\figsiz]{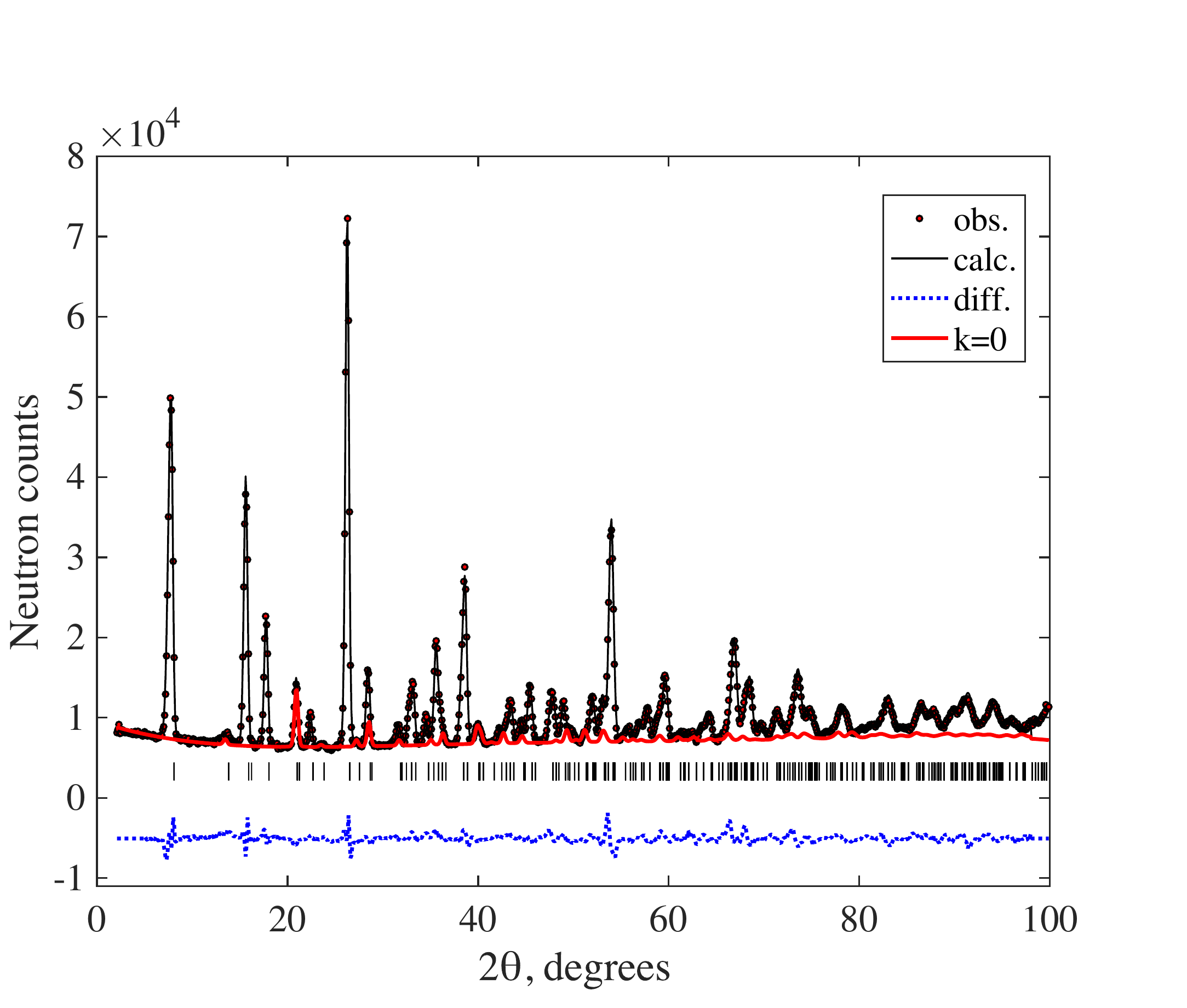} 
\par\end{centering}
\caption{ Neutron powder diffraction pattern showing the difference between data measured at $T=1.5$\,K and $T=30$\,K, with wavelength $\lambda=2.567$\,\AA. The solid line shows the result of the fit to the model with \Psix\ symmetry proposed here. The rows of vertical tick mark the hkl's of the Bragg peaks. The reliability factors are:   Rp:  3.48, Rwp:  4.56, Rexp:  1.30, Chi2:  12.3. The partial contribution of GM-point peaks is shown by the red curve.}
\label{new_P6_diff}
\end{figure}

\begin{figure}
\begin{centering}
\includegraphics[width=\figsiz]{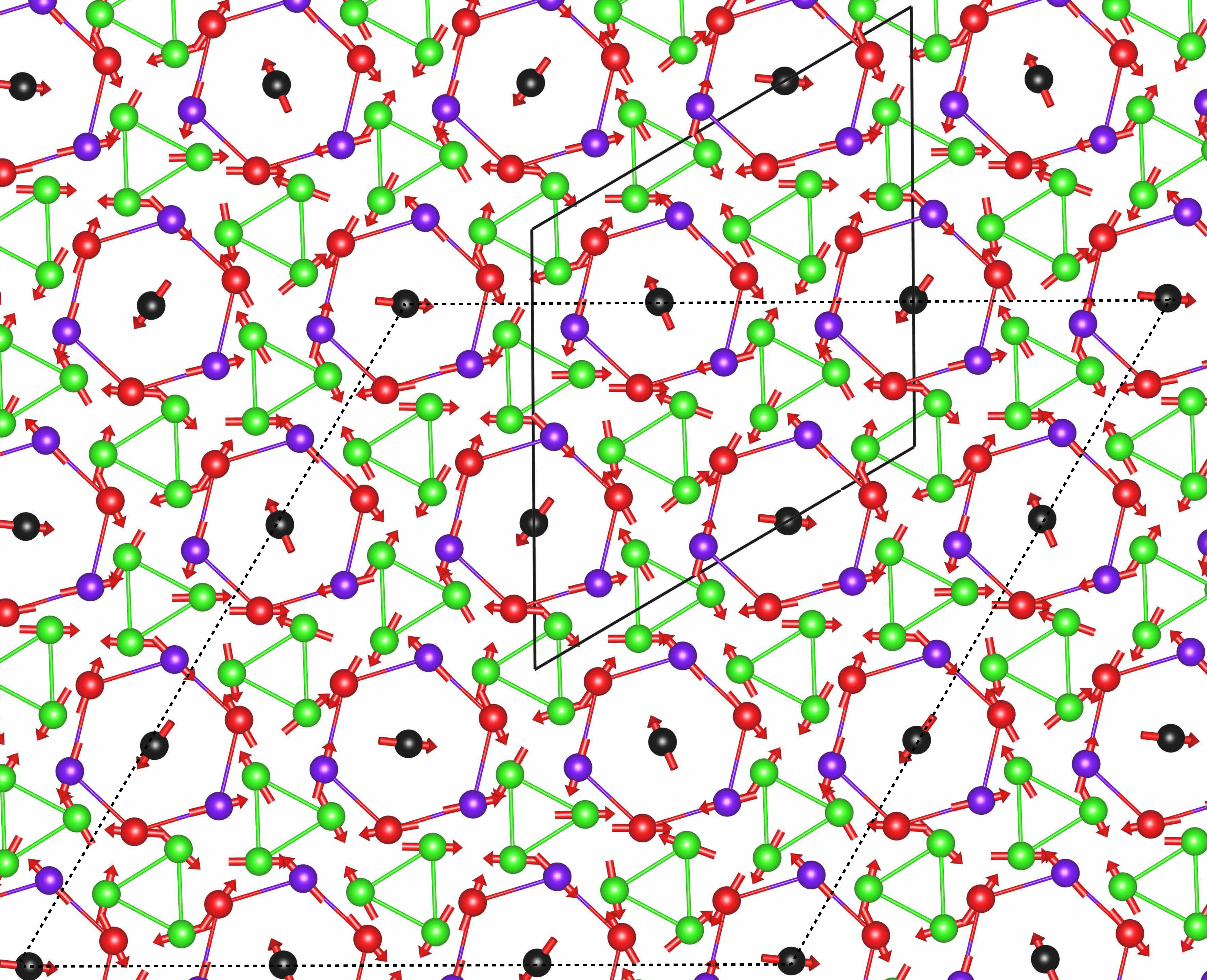} 
\includegraphics[width=0.5\figsiz]{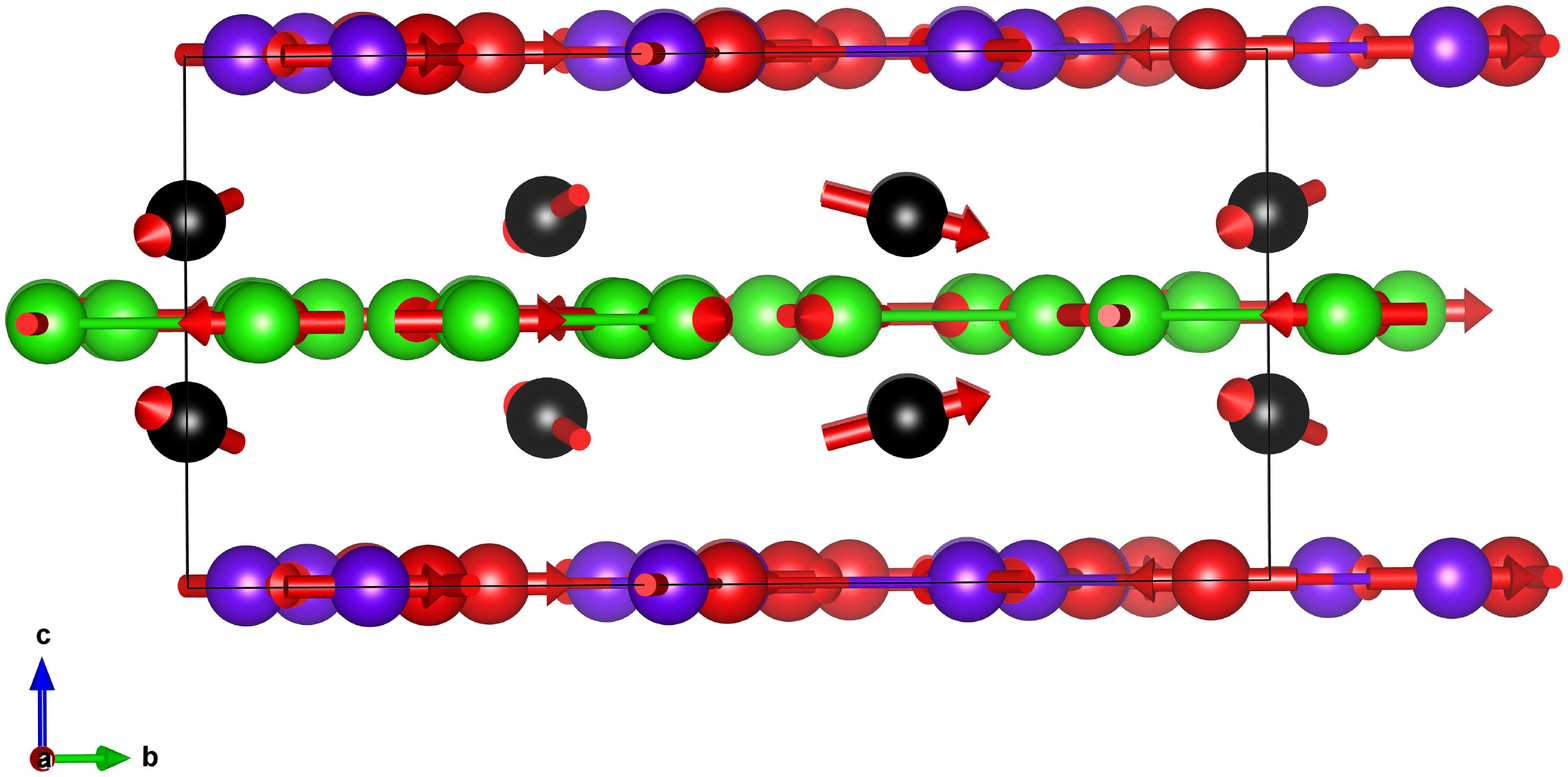} 
\par\end{centering}
\caption{View of the magnetic structure in projection on the xy-plane (top) and along the a-axis (bottom) corresponding to the refined model with \Psix\ symmetry. The unit cell is indicated by a black solid line. The dotted line shows a 3a x 3b supercell of the parent space group \P6m. Tb1 atoms are in green forming the triangles, Tb3 atoms are in black at the centres of hexagons. Tb2a\_1, Tb2a\_2 and Tb2a\_3  are in red, and the remaining three atoms derived from the Tb2 site are in blue.}
\label{new_P6} 
\end{figure}

The quality of fit for the structure model found in ~\cite{ISI:000232933700063} is shown in Fig.~\ref{oldK_k}. One can see that there are some misfits in comparison with the LeBail fit. The respective magnetic structure is shown in Fig.~\ref{old_Pm_irreps}. As mentioned above, three Tb2-moments break down the \Psix\ magnetic symmetry.  We first tried to force these moments to keep \Psix\ symmetry and have performed the structure fit. This fit was constrained in the same way as before, i.e. the magnetic moment magnitudes of the three atoms on each Tb2-orbit  were constrained to be equal. The quality of the fit is slightly worse, but comparable with the fit to the published structure. The reliability factors are Rp=8.48,     Rwp=12.6,  Rexp=1.31 and $\chi^2=92.2$, to be compared with the ones given in the caption to  Fig.~\ref{oldK_k}.
It is clear that we have to introduce the GM-point contribution that is allowed by symmetry. Using the structure parameters from the above fit  
as the starting values we have performed the fit under \Psix\ symmetry. The quality of the fit is illustrated in Fig. \ref{new_P6_diff}.  The solution in \Psix\ is well converged and is as good as the leBail fit, so there is no room for further improvement. We did the fit using spherical coordinates, assuming that all Tb moments have the same size. 
The angle for the first atom was fixed, because the structure factors are not sensitive to the overall phase shift. In other words the structure factors are not sensitive to the absolute orientation on the basal plane of the overall arrangement of moments. 
The magnetic structure is shown in Fig. \ref{new_P6}, where we have chosen the unit cell orientation similar to the Fig.~\ref{old_Pm_irreps} for comparison. The structure parameters are presented in the Table~\ref{str_tbl} and a CIF file, using the extended magnetic CIF format, with complete information about the structure, is included in the supplemental material. 

\begin{figure}
\includegraphics[width=0.5\figsiz]{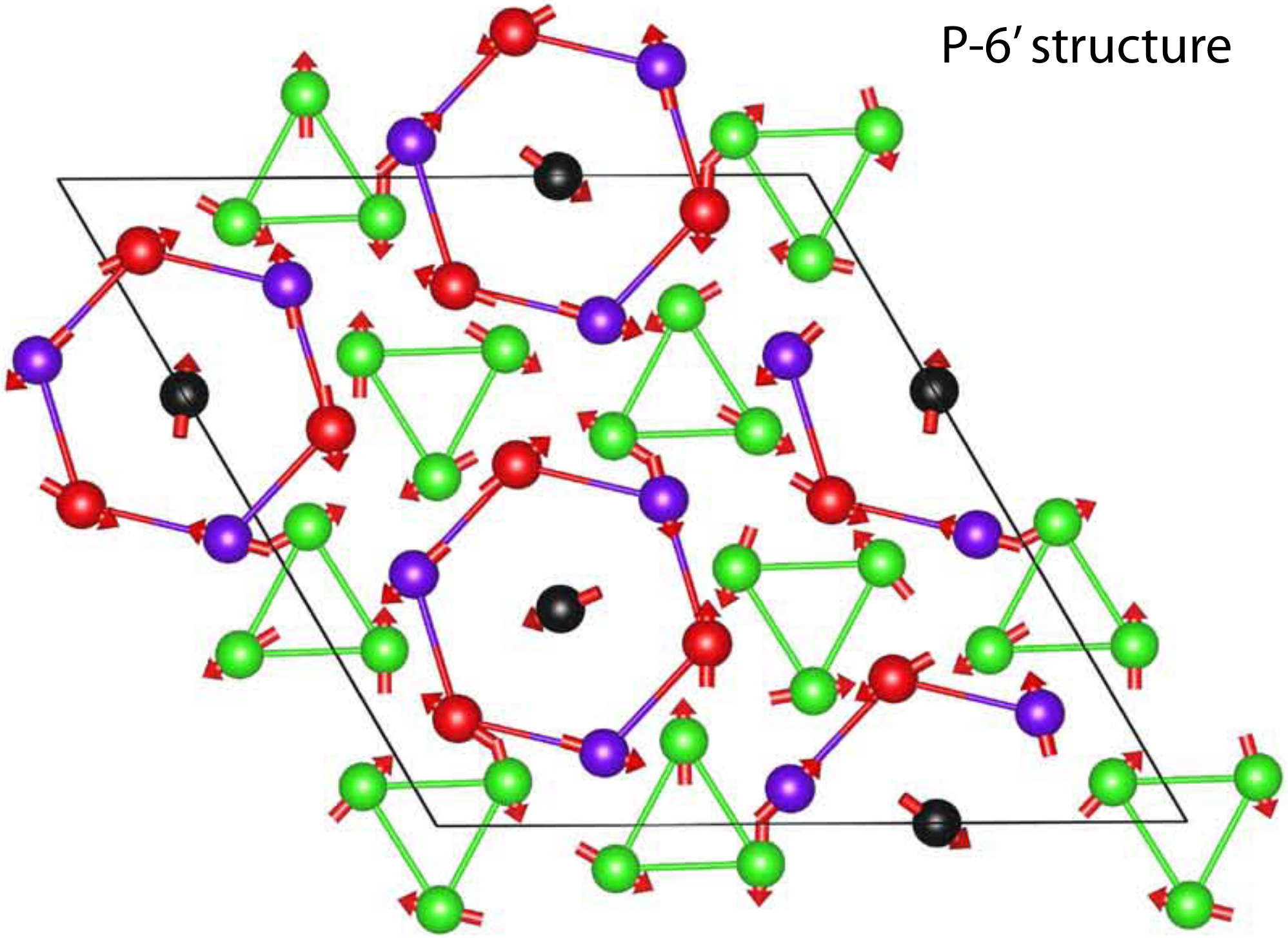} 
\includegraphics[width=0.5\figsiz]{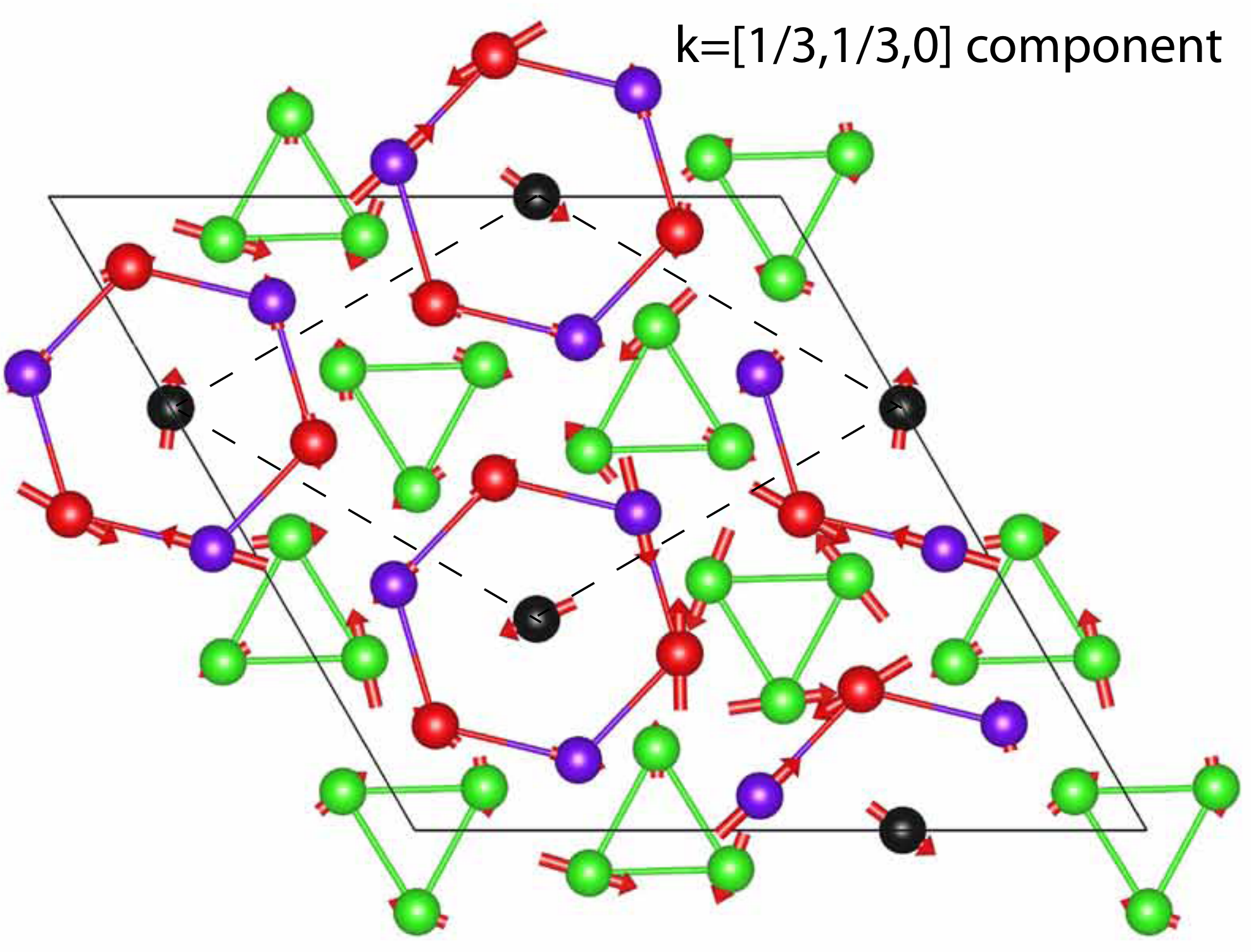} 
\includegraphics[width=0.5\figsiz]{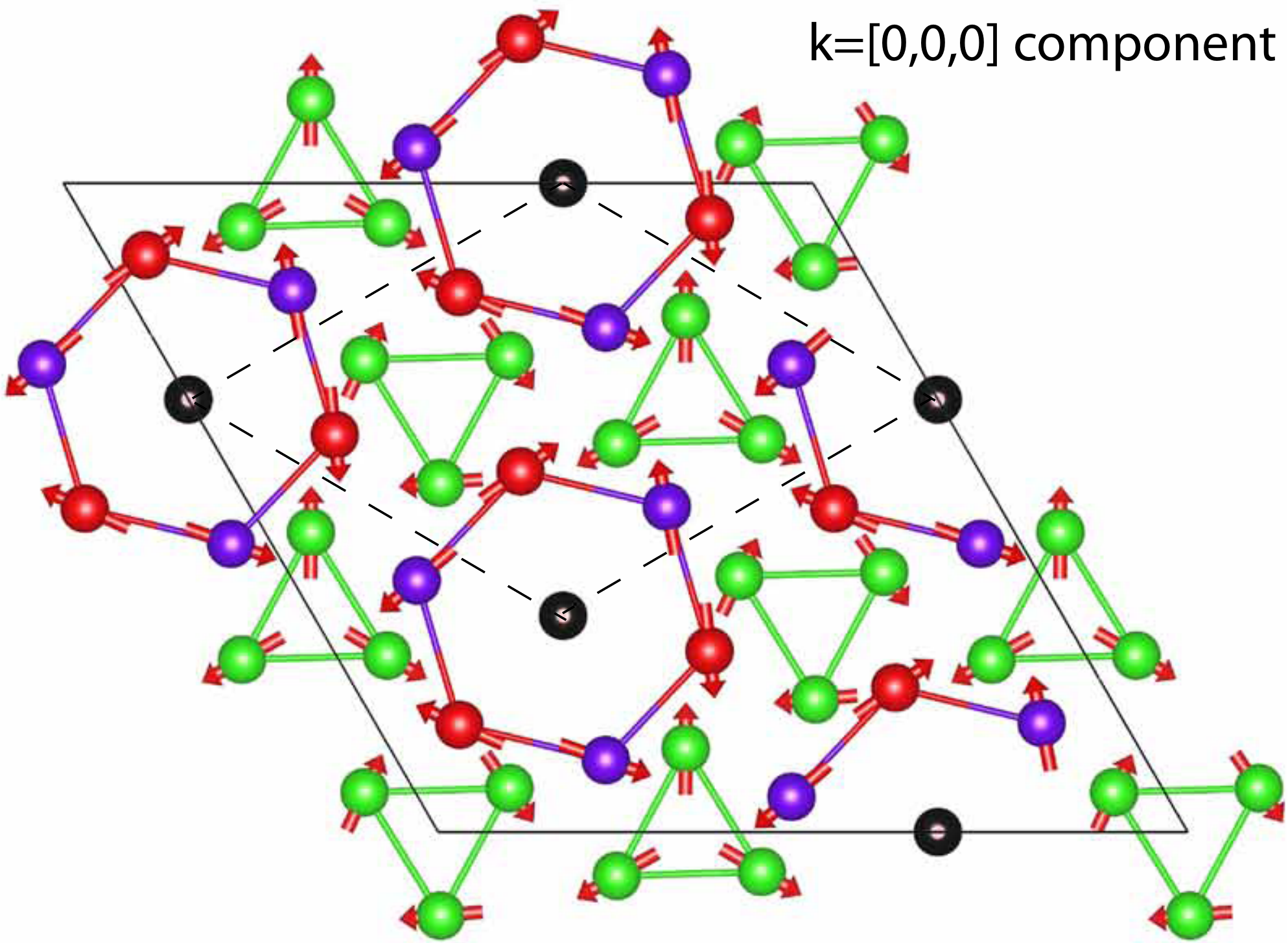} 
\caption{View of the magnetic structure in projection on the xy-plane corresponding to the refined model with \Psix\ symmetry (top), the partial contributions of magnetic modes with K-point \onek\ propagation vector (middle) and partial contribution of the magnetic modes with GM-point $[0,0,0]$ propagation vector (bottom). The GM-point contribution is smaller and has been scaled up by a factor of 3.333 in the plot. Tb1 atoms are in green forming the triangles, Tb3 atoms are in black being in the centres of hexagons. Tb2a\_1, Tb2a\_2 and Tb2a\_3  are in red, and the rest three atoms derived from the Tb2 site are in blue. The parent unit cell, in which the propagation vectors are defined, is shown by dashed lines.}
\label{new_P6_K_GM}
\end{figure}

It is interesting that the  GM-point contribution is rather large - it is separately shown in Fig.~ \ref{new_P6_diff}. 
This results in an overall modulation with the periodicity corresponding to the propagation vector $k_K$ but not purely sinusoidal. To further illustrate the partial contributions we have decomposed, using 
ISODISTORT\footnote {One needs to use Method 4: Mode decomposition of a distorted structure. First one loads the parent cif-file with \P6m\ structure, then - loads the mcif-file from the refinement in MSG \Psix, and follow the further steps to generate the separate mcif-files with modes for GM- and K-irreps.}
, the final solution into the symmetry modes of the GM- and K-points, which are shown separately in Fig.~\ref{new_P6_K_GM}. 
As expected, apart from the spin modulation associated with the irrep mK4K6, the spin arrangement also includes a significant $k_0$ contribution. It is remarkable that the two modes separately do not not maintain a common  moment magnitude for all atoms, and this is only attained through the combination of the two contributions. The GM component of the structure includes a mode associated with the irrep mGM1- for Tb3 with moments along the c-axis, and several modes according to irreps mGM1- and  mGM2+ for the moments on the basal plane of the Tb1 and Tb2 atoms. 

The difference diffraction pattern that was used to find and refine the magnetic structure contains purely magnetic scattering and is free of possible systematic uncertainties from the fit of crystal structure Bragg peaks, background, impurities, etc. So this was our preferable way for the magnetic structure determination.
However, to estimate the possible changes in the atomic positions in the magnetically ordered state we have undertaken a combined refinement of HRPT and DMC patterns at T=1.5~K. One cannot refine the crystal structure only from DMC data due to limited Q-range. For the crystal structure we used the average parent paramagnetic crystal structure model in space group \P6m, because the magnetic subgroup \Psix\ contains too many positional parameters. There are 13 Tb atoms and  38 Ag atoms with 113 parameters and the fit does not converge. 
The combined fit with the crystal structure in \P6m\ space group converged well with the atomic positions (19 parameters in total) within less than 1.5 standard deviation from their values in paramagnetic phase at 30~K for all parameters except four - 2.3 for x-AG1 and 1.9 for y-AG2, 2.0 for y-AG3, and 1.6 for y-AG4. We find these deviations insignificant. The detailed tables with the refined parameters are given in the supplementary information\footnote{Tables are in the file {SM\_pomjakushin\_Tb14Ag51.txt}}.  This type of fit gives considerably larger errorbars for the refined positional parameters and magnetic moment size, so the main purpose of such combined fit is to show that within the accuracy of our experimental data there are no significant changes in the crystal structure.


\section{Conclusions}

We have shown that the symmetry of the antiferromagnetic structure of \tbag\ with propagation vector \onek\  in the parent space group \P6m\ is given by one of the maximal symmetry groups that are possible for the multidimensional irrep mK4K6 that is active, namely an hexagonal MSG of type \Psix. This magnetic symmetry group constrains the possible mK4K6 ordering that can be present in the structure and it implicitly introduces third harmonic secondary degrees of freedom associated with propagation vector $k=0$, which demonstrate to be essential for an optimal fit of the experimental data. The structure consists of 13 independent Tb magnetic moments all having the same absolute moment value  $8.48(2)\mu_B$. For 12 atoms having m'.. site symmetry, the structure is a constant-moment purely cycloidal modulation in the ab-plane, while for one atom in a general position there is a substantial additional  helical contribution. The moment modulations are not completely sinusoidal, but have a significant third harmonic contribution, about 34\% in average amplitude.

\section{A{\lowercase{cknowledgements}}}

We acknowledge valuable contributions from A. Daoud-Aladine, A. Dommann and F. Hulliger.


\begin{table} 
\caption{The crystal and magnetic structure parameters of Tb atoms of \tbag\ in the refined model under magnetic space group \Psix. 
Tb1, Tb2 and Tb3 atoms are respectively in 3k, 3j and 6l Wykoff positions of this magnetic space group. The first two allow moment components only in (ab)-plane. The letters a and b in the atom name denote 1st and 2nd orbits following the notation used in Ref.~\cite{ISI:000232933700063}. The magnetic moments are given both in spherical coordinates $\theta, \phi$ and as components along the axes of the chosen hexagonal unit cell. The fit was done in spherical coordinates with the modulus of the magnetic moment constrained to be equal for all atoms, and a final refined value of $8.48(2)\mu_B$. The spherical coordinates are defined with respect to cartesian $x_c, y_c, z_c$ axes, which are chosen in the following way:  $x_c$-axis is along  the $a$-axis, the $z_c$-axis is along the $c$-axis, and the $y_c$-axis is perpendicular to the $(x_cz_c)$-plane. The moment components obtained from the spherical parameters are listed with non-significant digits in order to be consistent with the common moment magnitude. The unit cell, with \Psix in standard setting, is related with that of the paramagnetic phase in \cite{ISI:000232933700063} through the transformation (-1,1,0),(1,2,0),(0,0,1) and origin shift (1/3,2/3,0). The values of the lattice constants are $a=b,c $ 21.77944, 9.25767 \AA.}

\label{str_tbl}

\begin{tabular}{l l l } 
& $x \, y \,  z$  &  $\phi, \theta$, degrees \\ 
&           & $m_x\, m_y\, m_z,\, \mu_B$ \\

\hline 
Tb1a\_1  & 0.20453 -0.06757 0.5 & 330 90\\
 &     & 4.895 -4.895 0\\
Tb1a\_2  & 0.53787 0.5991 0.5 & 147(2) 90\\
 &     & -4.47 5.31 0\\
Tb1a\_3  & 0.8712 0.26577 0.5 & 211(3) 90\\
 &     & -9.79 -5.05 0\\
Tb1b\_1  & 0.79547 0.4009 0.5 & 130(3) 90\\
 &     & -1.63 7.55 0\\
Tb1b\_2  & 0.1288 0.06757 0.5 & 286(3) 90\\
 &     & -2.30 -9.39 0\\
Tb1b\_3  & 0.46213 0.73423 0.5 & 329(3) 90\\
 &     & 4.67 -5.12 0\\
Tb2a\_1  & 0.16813 0.61233 0 & 280(3) 90\\
 &     & -3.38 -9.65 0\\
Tb2a\_2  & 0.50147 0.279 0 & 89(3) 90\\
 &     & 5.00 9.79 0\\
Tb2a\_3  & 0.8348 -0.05433 0 & 265(3) 90\\
 &     & -5.56 -9.76 0\\
Tb2b\_1  & 0.83187 0.721 0 & 223(3) 90\\
 &     & -9.54 -6.70 0\\
Tb2b\_2  & 0.1652 0.38767 0 & 222(2) 90\\
 &     & -9.58 -6.55 0\\
Tb2b\_3  & 0.49853 0.05433 0 & 44(3) 90\\
 &     & 9.49 6.84 0\\
Tb3a\_1  & 0 0.66667 0.3104 & 84(2) 75(2)\\
 &     & 5.64 9.42 2.13\\

 \end{tabular}
  
\end{table}

\end{document}